# Strain-induced Moiré Reconstruction and Memorization in Two-Dimensional Materials without Twist

Nazmul Hasan[1], Tara Peña[1,2]*, Aditya Dey[3]*, Dongyoung Yoon[4], Zakaria Islam[4], Yue Zhang[4], Maria Vitoria Guimaraes Leal[1], Arend M. van der Zande[4,5,6], Hesam Askari[3], Stephen M. Wu[1,7†]

[1]Department of Electrical and Computer Engineering, University of Rochester, Rochester, NY, 14627, USA

[2]Department of Electrical Engineering, Stanford University, Stanford, CA 94305, USA

[3]Department of Mechanical Engineering, University of Rochester, Rochester, NY, 14627, USA

[4]Department of Mechanical Science and Engineering, University of Illinois, Urbana-Champaign, Urbana, IL, 61801, USA

[5]Department of Materials Science and Engineering, University of Illinois, Urbana-Champaign, Urbana, IL, 61801, USA

[6]Grainger College of Engineering, University of Illinois, Urbana-Champaign, Urbana, IL, 61801, USA

[7]Department of Physics and Astronomy, University of Rochester, Rochester, NY, 14627, USA

*These authors contributed equally to this work.

†stephen.wu@rochester.edu

## Abstract

Two-dimensional (2D) materials with a twist between layers exhibit a moiré interference pattern with larger periodicity than any of the constituent layer unit cells[1–3]. In these systems, a wealth of exotic phases appear that result from moiré-dependent many-body electron correlation effects or non-trivial band topology[4–17]. One problem with using twist to generate moiré interference has been the difficulty in creating high-quality, uniform, and repeatable samples[18] due to fabrication through mechanical stacking with viscoelastic stamps. Here we show, a new method to generate moiré interference through the controlled application of layer-by-layer strain (heterostrain) on non-twisted 2D materials, where moiré interference results from strain-induced lattice mismatch without twisting or stacking. Heterostrain generation is achieved by depositing stressed thin films onto 2D materials to apply large strains to the top layers while leaving layers further down less strained[19–23]. We achieve deterministic control of moiré periodicity and symmetry in non-twisted 2D multilayers and bilayers, with 97% yield, through varying stressor film force (film thickness × film stress) and geometry. Moiré reconstruction effects are memorized after the removal of the stressor layers. Control over the strain degree-of-freedom opens the door to a completely unexplored set of unrealized tunable moiré geometric symmetries[24–30], which may now be achieved in a high-yield and user-skill independent process taking only hours. This technique solves a long-standing throughput bottleneck in new moiré quantum materials discovery and opens the door to industrially-compatible manufacturing for 2D moiré-based electronic or optical devices[31–36].

**Introduction**

Two-dimensional (2D) moiré quantum materials have been a highly versatile platform for the discovery of novel correlated electronic or topologically non-trivial phases, which all result from the tunable electronic band structure caused by the moiré interference effect[15,16,30,37,38]. Effects such as superconductivity[4,13,14], correlated insulators[5,8], correlated magnetism[6,8], generalized Wigner crystallization[9,10], integer/fractional Chern insulators[11,17], and integer/fractional quantum anomalous Hall effects[7,12] have been found in a variety of twisted homo-bilayer 2D systems, or lattice-mismatched hetero-bilayer 2D systems. With the precise tunability of moiré interference within 2D van der Waals (vdW) bonded systems, it is possible to enable a generalized platform to discover newly predicted quantum phases or better understand the many-body physical principles that underpin existing ones[8,15,16,30].

However, a key challenge that has followed the moiré 2D materials class, from their initial discovery to the present day, is the difficulty in fabricating high-quality and repeatable samples of these systems[18]. The difficulty lies in the primary technique used to generate these samples, which relies on macroscale manual mechanical manipulation and stacking of individual layers using viscoelastic stamps. In this process, deterministically precise tuning of twist angle is difficult, as well as taming uncontrolled strains that are ubiquitous to every transfer. These unintentional and uncontrolled aspects of disorder in moiré materials have been well documented by the community[39–43], where ultimately a large amount of specialized user expertise and time are required to generate high quality samples. This process has been estimated to take on the order of hundreds of man-hours to create a single device in a low-yield process, which ultimately corresponds to multiple months of effort to generate the relevant materials for the experiment with limited repeatability. This represents a significant and fundamental bottleneck to the throughput of the discovery of new quantum materials or realizing the goal of achieving a universal quantum emulation platform[15,16,30]. Many researchers have attempted to solve this challenge through some form of specialized post-manipulation of twisted 2D materials[21,44–46], but none of these works introduce a universal method to fundamentally eliminate the largest uniformity bottleneck in fabrication, mechanical dry-transfer.

Here in this work, we solve this challenge by applying a completely different technique to generate moiré interference in 2D-bonded vdW systems, applying heterostrain (inequivalent strain applied to each 2D layer) to generate lattice-mismatch induced moiré reconstruction in non-twisted 2D systems. Essentially our technique allows us to create a lattice-mismatched heterostructure from a homostructure 2D system and eliminate the need for the van der Waals heteroepitaxial dry-transfer technique entirely. To achieve this, we apply our expertise in process-induced strain engineering on 2D materials[19–23,47–50]. Stressed thin films are deposited onto 2D materials (Fig. 1a), where film stress relaxes by film expansion or contraction. The underlying 2D material is strained with this relaxation process, where the strain applied to the top layer of exfoliated 2D flakes is greater than strain applied to each subsequent layer due to weak vdW interlayer bonding. Moiré reconstruction occurs after exceeding the heterostrain threshold for strain soliton formation. When the stressor layer is removed, through a wet etching technique, the moiré reconstruction effects are memorized (Fig. 1a) and can be directly imaged[43,51–53]. Process-induced strain engineering is also known to enable the design of arbitrary 2D strain profiles through the choice of different lithographically patterned geometric shapes[23,54]. In this work, we also demonstrate the utilization of stressor geometry design to enable arbitrary periodicity and symmetry moiré interference effects, which go beyond just twist-alone and can only be achieved with the designed control of the strain degree-of-freedom.

There exists a generalized continuum of moiré interference patterns achievable with the engineered control of the 2D strain tensor in bilayer 2D materials[24,25]. Demonstrating a deterministic, high-yield method to generate these new moiré symmetries with tunable moiré periodicity, would represent a fundamental shift in the engineered design of moiré quantum materials and a solution to an otherwise difficult throughput bottleneck of moiré fabrication. Additionally, process-induced strain engineering[55,56] and the stress memorization technique (SMT)[55,57–60] have been used extensively in industrial practice for Si-based transistors since the 90 nm technology node in 2003. Due to this long industrial history, we anticipate our strain-defined moiré engineering and memorization technique may allow for the translation of the positive benefits of robust, high-yield, wafer-scale, nanoscale CMOS manufacturing to the world of 2D quantum materials. Beyond fundamental physics discoveries in heterostrained moiré systems, we also envision this approach to have strong impact in applied technologies, enabling the manufacturable design of 2D sliding ferroelectric devices or other related electronic, optical and quantum electronic technologies utilizing moiré nanoengineering in 2D systems[31–36].

**Moiré Reconstruction from Process-induced Strain Engineering**

Figure 1a illustrates how heterostrain induced moiré reconstruction is achieved in 2D vdW materials without rotational twist. 3R- and 2H-$MoS_2$ flakes are mechanically exfoliated on Si/$SiO_2$ substrates with bilayer to bulk thickness. The exfoliated samples are uniformly encapsulated with a stressed thin film through e-beam evaporation (details of films in Methods), where the film stress is determined by the type of material deposited[61]. Biaxial heterostrain is generated directly proportional to the film force (film stress × film thickness)[19], quantified in Supplementary Fig. S1. For a large enough magnitude heterostrain, $MoS_2$ develops a strain-induced lattice mismatch between layers resulting in moiré reconstruction. The blanket stressor can subsequently be removed from the samples with a material selective wet etching process, to etch away the stressor layers while ensuring the 2D flakes are not damaged. This etch-based unloading process allows for the top surface to be exposed, where any memorized moiré reconstruction effects may be subsequently visualized using tilt-angle dependent scanning electron microscopy (SEM)[43,51–53]. Optical micrographs of the three steps during the process are presented in Fig. 1b. The selected flakes are re-imaged after each step using an optical microscope to confirm high-quality encapsulation and etching. To visualize this reconstruction process, 3R-MoS$_2$ flakes are imaged after tilting the SEM stage at an 18º angle relative to the electron beam, allowing the observation of any AB/BA domain reconstruction contrast differences[43,62]. More details of the fabrication process and imaging techniques are included in the Methods section.

In Fig. 1c, the image of an as-exfoliated non-twisted 3R-MoS$_2$ flake is shown, which shows no signature of moiré reconstruction with a flat surface with no apparent distortions. After applying and removing the stressor through etching, the same sample is reimaged to directly visualize the stressor effect on the flakes. The post-encapsulation 3R-MoS$_2$ shows AB/BA domain reconstruction patterns induced by the straining process, exhibiting patterns consistent with strain soliton network formation with regular triangular lattice reconstruction at the smallest periodicity lengthscales. This heterostrain-induced soliton network formation process is reproducibly observed across all investigated samples. Stacking-sensitive contrast in SEM imaging also reveals the resulting reconstruction patterns of both the 2H phase (H-type reconstruction, hexagonal domains) and the 3R phase (R-type reconstruction, triangular domains) of MoS$_2$ (Fig. 1d). This demonstrates that heterostrain is capable of generating lattice specific soliton network domain reconstruction that mimics both near-0° twisted MoS$_2$ bilayers (R-type), as well as near 60° twist (H-type)[63,64].

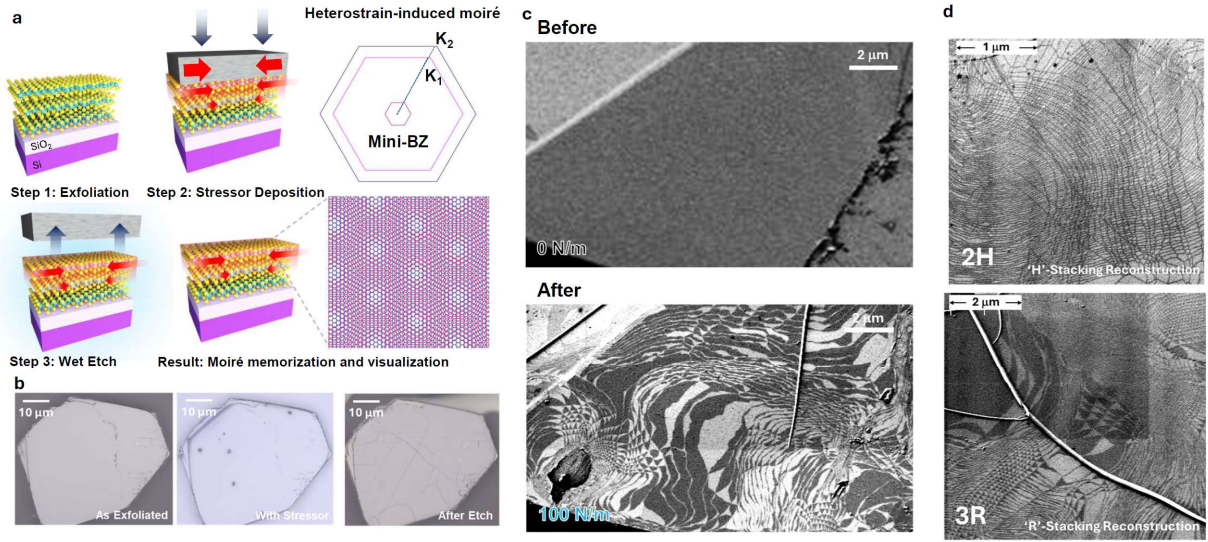

**Figure 1:** (a) Depiction of process-induced heterostrain engineering mechanism resulting in moiré memorization effect. Real and momentum space representation of heterostrain induced moiré interference, including top and bottom layer Brillouin zone and subsequent mini-Brillouin zone. (b) Optical micrograph of typical exfoliated flake, as-exfoliated, with deposited stressor layer, and after stressor etching. (c) Tilt-angle dependent scanning electron micrograph of 3R-$MoS_2$ reconstruction before and after 100 N/m stressor application. (d) Reconstruction visualized for both 2H-$MoS_2$ and 3R-$MoS_2$ after 160 N/m and 140 N/m stressor application respectively.

## Molecular Statics Simulation of Strain-induced Moiré Formation and Memorization

To understand the heterostrain-induced moiré reconstruction effect better, we performed atomistic simulations using the Molecular Statics (MS) method to investigate the atomic-scale mechanisms governing moiré formation. The deposited stressor in experiments demonstrates strong adhesion to the top layer of the exfoliated 2D material, which is then weakly transmitted to the underlying layers. In our model, we assume similar high top-layer adhesion, straining only the topmost layer. This applied strain is subsequently transmitted to the R-stacked $MoS_2$ layers below through interatomic forces governed by our Machine Learned Interatomic Potential (MLIP)[65,66], detailed in Supplementary Fig. S2. No additional adhesion or boundary conditions are defined between the layers, and the deformation in the underlying layers arises purely from strain transfer via interlayer vdW interactions. We present the effective strain maps in Fig. 2a for the second and third layers, representing $1^{st}/2^{nd}$ layer and $2^{nd}/3^{rd}$ layer reconstruction, respectively. We also include a $4^{th}$ layer in MS simulation to account for complete interactions of the $3^{rd}$ layer, but a negligible strain transfer is shown in this layer due to the weakness of the vdW interactions. The effective applied strain in the first layer is calculated as $\varepsilon_b = \sqrt{\varepsilon_{xx}^2 + \varepsilon_{yy}^2 - \varepsilon_{xx}\varepsilon_{yy} + \varepsilon_{xy}^2}$, where $\varepsilon_{xx}$, $\varepsilon_{yy}$, and $\varepsilon_{xy}$ are the normal and shear strain components, respectively[67]. To enhance spatial resolution and capture localized features, each strain map is normalized relative to the magnitude of the applied top-layer strain.

The strain profiles across the layers reveal three distinct deformation regimes delineated by two interlayer slip transitions. As shown in Fig. 2b, the initial regime (labeled Stage-I, up to ~1.6% strain) is characterized by a linear and monotonic strain transfer, driven by the elastic coupling across the interface with no slippage. However, beyond this point, a notable reduction in the strain transfer efficiency emerges, suggesting the onset of interfacial weakening. This defines an intermediate regime of partial interlayer slippage (Stage-II, between 1.6 and 2.35% strain) where vdW interactions begin to deteriorate under increased loading. Beyond this stage the strain transfer experiences a sharp decline. At this point, the system enters a major slip regime (Stage-III, beyond 2.35% strain), characterized by significant interfacial decoupling and a loss of shear-bearing capacity.

The three-stage slippage process is closely tied to the evolution of strain localization and pattern formation in the underlying stacked layers. In the initial no-slip stage, the top layer efficiently transmits strain to the second layer, leading to the development of channel-like regions (white regions in contour plots, Fig. 2a). As strain increases these regions grow and merge indicating the onset of partial slip, visualized as blue zones in the strain contour map. Continued loading further transforms these patterns, ultimately giving rise to well-defined triangular moiré domains as full slippage is achieved. Strain localization in the third layer follows a similar progression, governed by the strain transferred from the second layer during loading.

We examine the stability of the simulated moiré sub-domains by gradually releasing the applied strain until the film experiences zero net force at its edges. This unloading process is conducted separately for each of the three slippage stages, as illustrated in Fig. 2c. In the no-slip stage, as the strain is removed, the sub-domains progressively fade, and the film reverts to its original starting configuration with no detectable residual strain, indicating elastic recovery. In contrast, unloading from the partial and full slip stages yields more complex responses characteristic of plastic deformation, with portions of the strain retained within the film even after complete unloading. During partial slip, the skewed triangular domains relax into spiral-shaped patterns. Unloading from the full slip stage, on the other hand, preserves the well-formed triangular moiré domains with sharp edges. Despite the removal of external stress, both the moiré structures and their associated strain distributions remain unchanged, showing no configurational alteration. Figure 2d presents the average film force versus strain during both loading and unloading. For the no-slip case, the force-strain curve retraces its loading path back to zero force, confirming elastic behavior. However, the curves for partial and full slip unloading deviate from this path, resulting in residual strain. These unloading results clearly highlight plastic deformation behavior, leading to memorized reconstruction of moiré patterns into spiral and triangular domains, respectively.

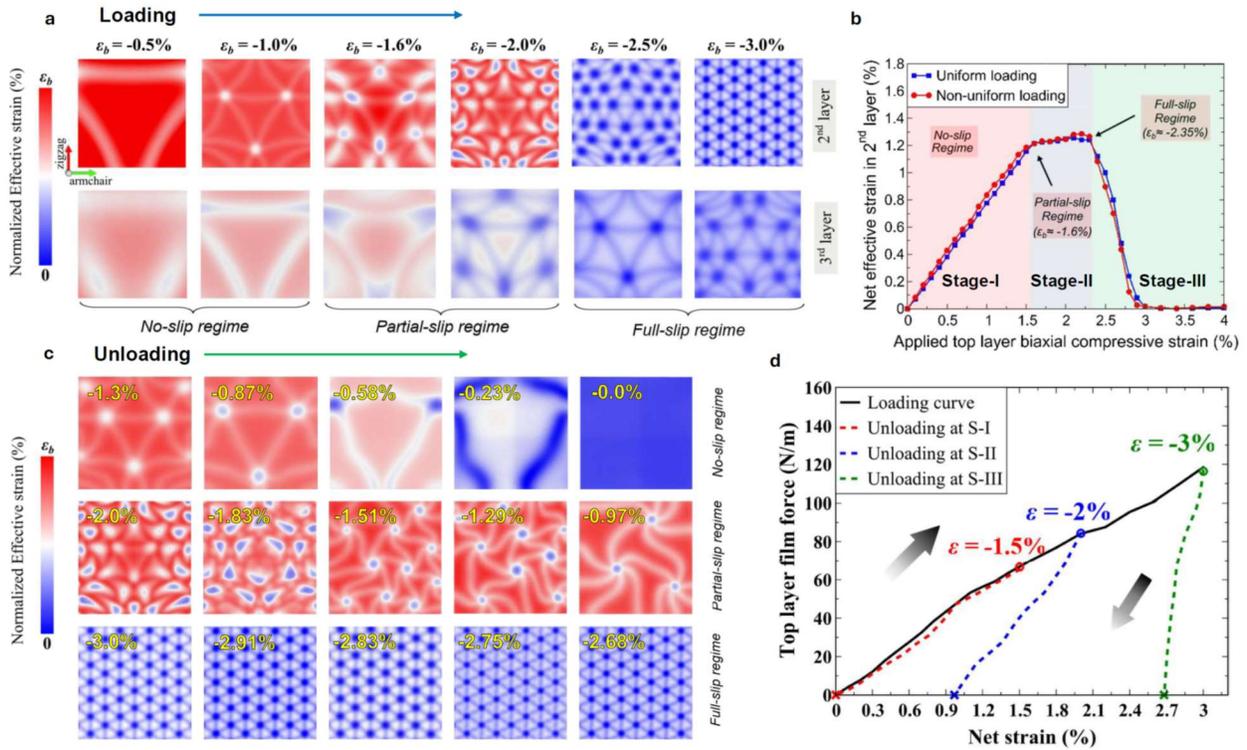

**Figure 2:** (a) Effective strain distribution contour plots of second and third layer of the 3R-MoS$_2$ atomistic model while applying biaxial compression to the top layer. (b) The average strain in the layer beneath the top layer is plotted as a function of the applied strain in the top layer, indicating three regimes (Stage-I, no-slip, Stage-II, partial-slip, Stage-III, full-slip). (c) Strain memorization effect displayed as effective strain contour plots upon unloading strain to the top layer from three different stages. (d) The loading–unloading curve is shown as a plot of film force versus net residual strain in the top layer.

**Strain Magnitude Dependent Moiré**

To directly test the strain magnitude dependent loading and unloading behavior from simulations, we examine samples prepared with increasing film force, since film force is directly proportional to strain applied to the top 2D layer[19]. Shown in the first image Fig. 3a, and labeled 0 N/m, is the typical natural behavior of as-exfoliated 3R-$MoS_2$ before the application of stressor layers, with minimal domain reconstruction present. Next, stressors are applied with increasing film force and etched away to mimic the loading and unloading process. At 40 N/m the flakes appear with AB/BA stacking reconstruction, indicating the emergence of strain soliton networks with exceptionally large length scales (>10 μm) and wavy domain walls. As the applied film force increases, large spiral domains evolve into denser structures with reduced length scales and periodicity, exhibiting increasingly complex patterns. By 100 N/m, these domains begin to transform into regular triangular lattice AB/BA stacking configurations, consistent with expectations from simulation when we approach the full slippage condition. Eventually, at 120 N/m we observe uniform, periodic triangular lattice moiré reconstruction patterns, consistent with surpassing the partial slip threshold. At even higher film force of 140 N/m, highly regular features emerge with uniform equilateral triangles with smaller periodicity than at lower film forces.

The local reconstruction-induced domain density is quantified for a large set of samples (Fig. 3b), allowing us to view the effect of strain on domain density at each film force application. We separately quantify domain density for regions with the maximally dense uniform triangular domain reconstruction per flake, and for regions with the maximally dense non-uniform wavy domain reconstruction per flake. The trend initially exhibits close to zero domain density at <80 N/m, consistent with elastic regime behavior, but begins to rise when we reach the 80 N/m threshold, consistent with the elastic and partial slippage threshold. As film force is further increased, within the data for fully uniform triangular reconstruction, we observe a quadratic dependence of domain density with film force. We match these results with theoretically calculated domain density from simulations (see Methods), where we see good correspondence with theory for the highest domain density values at each film force. This result also reveals that our SEM based visualization is showing the reconstruction process from the $2^{nd}/3^{rd}$ layer, as the theoretically predicted values of $1^{st}/2^{nd}$ layer reconstruction effects exceeds the limit of detection in the SEM. Additional quantification of the $1^{st}/2^{nd}$ layer reconstruction is presented in supplementary Fig. S3, where torsional force microscopy (TFM) is used to resolve the smaller lengthscale moiré reconstruction effects. Similar triangular AB/BA domain reconstruction behavior is observed compared to the SEM-based measurement presented in Fig. 3a, albeit at a smaller scale, as expected from simulation. Overall reconstruction patterns as measured with TFM mimic the behavior observed in our SEM measured samples, and therefore we focus on the higher-throughput SEM-based evaluation method to first understand the underlying mechanical principles behind heterostrain-induced reconstruction.

When examining the domain density for non-uniform reconstruction regions on the samples, there appears to be a limit to how dense these non-uniform regions may become. This effect may relate to the fact that large strain needs to be applied before we reach the full slippage regime, and in non-ideal cases there may be regions that do not achieve the maximal ideal strain due to non-ideal strain transfer behavior between layers. This may occur due to fracture/slip of the interface between the stressor and the 2D material, leading to premature unloading. Premature unloading may lead to release of the sample in the partial slippage regime leading to increasingly dense non-uniform regions that approach a threshold of ~75 $\mu m^{-2}$. The non-uniform patterns formed in our samples match the expected simulated behaviors of domain reconstruction under non-uniform loading (Supplementary Fig. S4). We later solve this limitation by changing the stressor boundary condition through lithographic patterning, described in later sections.

To further demonstrate that the effects we observe are due to heterostrain alone, we quantitatively examine the waviness of the domain walls, as we transition from the partial slippage regime to the full slippage regime. In simulations it is shown that domain walls between reconstructed regions transition from spiral to triangular reconstruction as the characteristic moiré reconstruction scale is made smaller by larger strains. In Fig. 3c, this effect is quantified by the percent difference in node-node domain wall length compared to a straight line drawn between two nodes. This effect is plotted as a function of node-to-node distance across seven different samples, where within each sample we have a wide variation of node-to-node spacing. We observe a universal scaling effect where, across all domains and within all samples, the maximum value of waviness is reduced as node-to-node distance is decreased. This waviness parameter eventually reduces to zero at a characteristic node-to-node spacing of ~100 nanometers. This matches the expectations from simulations, and is a unique effect from heterostrain-induced moiré reconstruction.

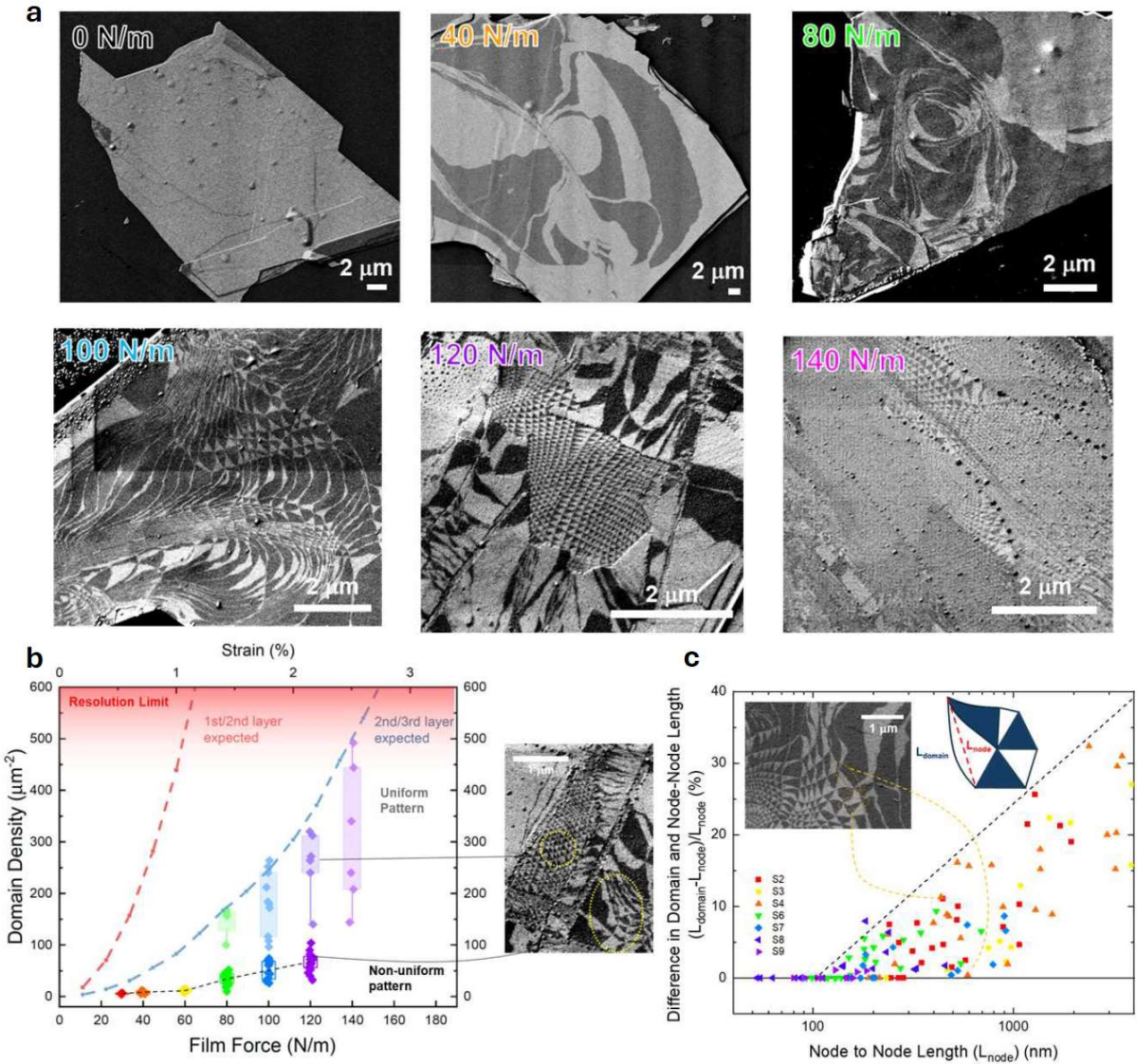

**Figure 3:** (a) Representative results of reconstruction from samples after applied stressor layers (0-140 N/m) and removal. Reconstruction begins with spiral domains, which transition to triangular domains. (b) Quantified domain density for multiple samples in both uniform triangular reconstruction regions on each sample, and in non-uniform reconstruction regions, as a function of applied film force. Values are compared to theoretically calculated domain density from 1st/2nd layer reconstruction and 2nd/3rd layer reconstruction. (c) Quantified domain waviness as a function of node-to-node spacing, depicting a natural transition from wavy domain walls from spiral reconstruction to straight domain walls from triangular reconstruction, as domain density increases.

**Designable reconstruction with patterned heterostrain profiles**

To design the heterostrain-induced reconstruction effect and to increase determinism/yield of our process, we explored using patterned stressor layers on bilayer and multilayer 2D materials. Figure 4a presents an optical image of a square pattern $MgO_x$ stressor (54×54 µm, 30 nm thick, -30 N/m), fabricated with shadow mask lithography, on a 2H bilayer $MoS_2$ on a $Si/SiO_2$ substrate (see Methods). We use hyperspectral Raman spectroscopic mapping to non-invasively probe the influence of the heterostrain through the transparent $MgO_x$ stressor. Inside the stressor, the $E^1_{2g}$ and $A_{1g}$ modes blue-shift, consistent with previous results showing that $MoS_2$ under $MgO_x$ stressor experiences tensile strain[19,23]. Figure 4b and 4c respectively show maps of the $E^1_{2g}$ and $A_{1g}$ mode wavenumber for the same region shown in the optical image in Fig. 4a. In the Raman map discrete jumps in Raman peak position show up as complex networks of lines in the mode positions, with straight lines near the edges and disordered lines near the corners. At the center of the stressor the mode positions form a relatively uniform patch. In comparison, monolayer $MoS_2$ and heterobilayer $MoS_2/WSe_2$ on the same kind of $SiO_2$ substrate both show a smoothly varying mode position underneath the stressor with higher strain concentrated near the edges[23]. The presence of two discrete steps in the Raman mode wavenumbers, along with distinct behavior within each step, suggest the existence of two separate strain zones, elastic (Stage-I) at the center and transitioning to partial slip (Stage-II) and full slip (Stage-III) at the edges where strain magnitude is higher. We interpret the discrete jumps in the 2H bilayer $MoS_2$ as a signature of heterostrain induced slip creating solitons and changes in stacking order just like those observed in the multilayer samples seen in SEM images (Fig. 1 and 3). Unfortunately, there are no well-defined Grüneisen parameters for bilayer $MoS_2$ that account for stacking faults, so it is not possible to quantify the strain, and we must rely on a qualitative description of the behavior. Supplementary Fig. S5-S6 provides additional analysis of the Raman spectroscopic maps and individual spectra.

Next, we examine the effect of patterned and removable stressors on multilayer 3R-MoS$_2$. This represents the lithographically patterned variant of the previously presented uniformly encapsulated results from Fig. 1 and Fig. 3 (see Methods for more detail). In Fig. 4d, optical images are shown for a selectively applied square stressor on an as-exfoliated 3R-MoS$_2$ multilayer flake, before, during and after the stressor application and etch process. Figure 4e presents a visualization of the domain reconstruction effects after the removal of the stressor in the lithographically defined region after etch, and Figure 4f shows a zoomed in view near one corner. Similar to the results from Raman spectroscopic mapping, but with finer detail due to the resolution of the SEM, we see unique geometrically defined strain soliton networks. At the edges, we observe linear soliton network formation, with increasing density closer to the edge. Near the corner we observe triangular soliton network formation, with increasing density closer to the corner. Near the center of the square, we see large area wavy domain reconstruction behavior. To interpret these results, Figure 4g-j show finite element analysis (FEA) of the (g) maximum principle strain and (h-j) the strain components in a 2D monolayer allowing slip with respect to the bulk, similar to previous works on patterned stressors[23,47,54]. These maps reveal that near the center of the square, the strain is small and biaxial, near the edges there is a large strain anisotropy and near the corners there is a large shear. From the finite element analysis, we interpret the differences in the soliton networks as rising from the different strain magnitude and type. Near the center, the small biaxial strain leads to a low soliton density without clear direction due to regions of elastic or partial slippage. Near the edges, the large anisotropic strain leads to linear soliton networks, similar to what is observed in materials under uniaxial strain[67]. Near the edges, the large shear component leads to the different symmetry of triangular domains in the soliton network. These effects in total represent an example of the deterministic creation of heterostrain induced moiré reconstruction that is determined by the geometric design of the stressor layer, and thus 2D strain tensor profile.

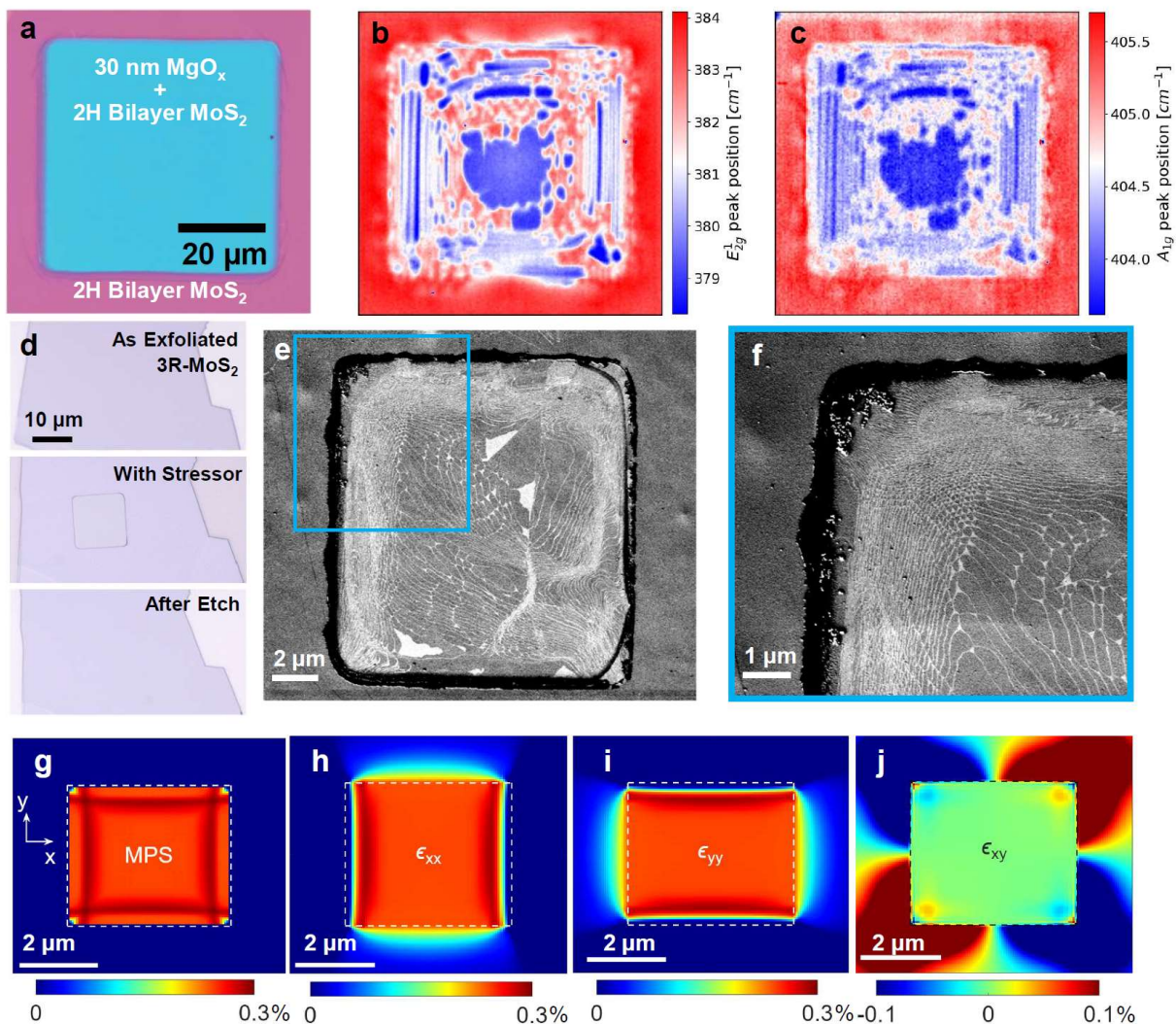

**Figure 4:** (a) Optical image of a patterned -30 N/m square $MgO_x$ stressor on 2H-$MoS_2$. (b) $E^1_{2g}$ peak position, and (c) $A_{1g}$ peak position from a Raman hyperspectral map of the bilayer presented in (a). (d) Optical images of a 3R-$MoS_2$ multilayer flake, before, during, and after the application of a lithographically patterned 44 N/m square removable Al/Au/Cr trilayer stressor. (e) Tilted SEM imaging of AB/BA reconstruction in the region after square stressor etching, with a zoomed view of the corner presented in (f). (g-j) Finite Element Analysis maps for a square stressor of the (g) maximum principal strain (MPS) (%), (h) strain component ($\epsilon_{xx}$) (%), (i) strain component ($\epsilon_{yy}$) (%), and strain component ($\epsilon_{xy}$) (%). To make the strain variation inside the stressors visible, the color scales in the FEA maps are intentionally limited to not showing the full range of possible strains outside the edges of the stressors.

## Symmetry, Determinism and Yield

Beyond just square stressor geometry, we may explore the effects of other stressor geometries as well. In Fig. 5, we present SEM images of domain reconstruction in 3R-$MoS_2$ under applied triangular, hexagonal, and elliptically patterned stressor layers. In all cases presented in Fig. 5, near straight edges the strain applied is more uniaxial in nature with larger strain perpendicular to the edge, resulting in linear strain soliton networks parallel to the edge. Near corners, the strain applied is more biaxial and/or shear in nature, leading to triangular lattice strain soliton networks resembling moiré reconstruction. In all cases, the reconstruction density increases as it nears the boundary of the stressor, due to the increase in strain magnitude as the edge is approached in lithographically patterned stressors[23,56]. In Fig. 5b,d,f the periodicity of the moiré reconstruction effect becomes unresolvable as it reaches the resolution limit of the SEM (<50 nm). It is also clear from Fig. 5b,d,f that the reconstruction patterns are not only linear or only equilaterally triangular, but represents a geometrically definable gradient between these two regimes through the design of the stressor geometry. These results match expectations from FEA simulations of strain patterns generated by the three geometric shapes presented in Supplementary Fig. S7. Supplementary Fig. S8 highlights the reconstruction effects on geometrically patterned stressors on 2H-$MoS_2$.

Finally, we quantify the determinism and yield for the heterostrain-defined reconstruction process. Presented in Supplementary Fig. S9 is a typical example of eight square pattern stressors applied to 3R-$MoS_2$. From this data, it becomes clear that our geometrically applied stressors perform the best when there are no pre-existing deformations on the 2D crystal to begin with, such as layer thickness changes, bumps, folds, or cracks. Deformations serve as alternative nucleation points for strain soliton formation causing deviation from the designed reconstruction pattern. Under ideal conditions without these deformations (32 total samples) we observe that 97% display triangular lattice reconstruction in at least one corner in each pattern, and out of all corners in total 68% display triangular lattice reconstruction. When accounting for all total samples fabricated, with and without deformations (82 total samples), we observe 91%, and 49% in the same categories listed above, respectively. Future studies in boundary conditions, stressor adhesion, stressor-2D fracture limits, and compatibility with wafer-scale grown 2D bilayers may further refine this process. This yield represents a significant advance in moiré fabrication throughput since this technique allows for arbitrarily large numbers of lithographically defined stressors to be patterned per chip (20-30 at a time in our demonstrations). Devices were fabricated on the order of hours, and ultimately always resulted in high numbers of successfully reconstructed flakes each time the technique was repeated.

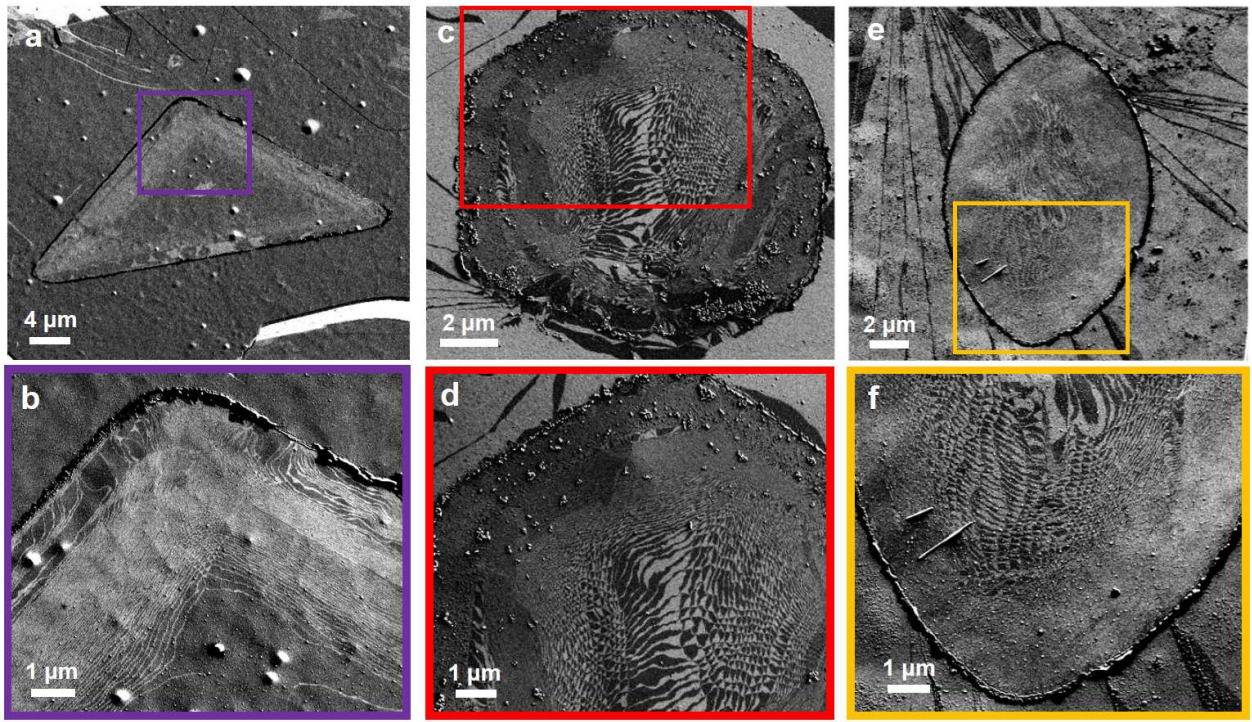

**Figure 5:** (a-b) Reconstruction for a 3R-$MoS_2$ sample with triangular stressor geometry with a 60 N/m stressor application and removal from wide (a) and zoomed (b) view. (c-d) Reconstruction for a 3R-$MoS_2$ sample with hexagonal stressor geometry with a 44 N/m stressor application and removal from wide (c) and zoomed (d) view. (e-f) Reconstruction for a 3R-$MoS_2$ sample with elliptical stressor geometry with a 50 N/m stressor application and removal from wide (e) and zoomed (f) view.

## Conclusion

We have presented a simple process-compatible method to engineer moiré reconstruction by design. Using process-induced strain engineering methods that have industrial-scale compatibility and robustness, we have shown that heterostrain induces layer-dependent lattice mismatches that effectively allow us to craft arbitrary 2D moiré heterostructures out of as-exfoliated homostructures. These effects work down to the bilayer level, and arbitrary strain patterns may be designed through stressor lithographic patterning[23]. This may allow for new moiré symmetries, achievable only through strain engineered design[25]. Analysis of the uniformity of the reconstruction effect, within regions that exhibit full-slippage, reveal that the variation in moiré domain size is equivalent to strain variations in the ±0.08% range, presented in supplementary Fig. S10. This is comparable to values of the best uniformity achieved[45], over similarly large areas, in mechanically stacked twisted moiré heterostructures. Through further refinement of all techniques presented, we anticipate a universally applicable, high-throughput method, to create deterministic moiré quantum materials by design with no specialized user skill involved. These techniques provide a pathway to break the key fabrication bottleneck of achieving high-throughput designer 2D moiré quantum materials, bringing along all the time-tested benefits of industrial-scale reliability from conventional semiconductor manufacturing.

## Acknowledgements

We wish to acknowledge support from the National Science Foundation (OMA-1936250 and ECCS-1942815). T.P. acknowledges the NSF MPS-Ascend postdoctoral fellowship. D.Y, Z.I., Y.Z, and A.M.v.d.Z. were supported by the National Science Foundation through the University of Illinois at Urbana-Champaign Materials Research Science and Engineering Center DMR-2309037. The authors acknowledge the use of facilities and instrumentation supported by NSF through the University of Illinois at Urbana-Champaign Materials Research Science and Engineering Center DMR-2309037. Torsional force microscopy was conducted at Stanford Nano Shared Facilities (SNSF), supported by the National Science Foundation, award ECCS-2026822. Author Nazmul Hasan dedicates this work and its outcomes to the memory of his beloved daughter, Wajiha Samawat Junaisha, who tragically passed away at 20 days old due to congenital heart defects during the first year of his graduate research journey.

**Methods**

**Sample Preparation and Fabrication**

We begin with 90 nm $SiO_2$ on Si substrates, which are sequentially cleaned in acetone and isopropanol using an ultrasonic bath for 15 minutes each, followed by nitrogen drying. The substrates are immediately transferred to an MBraun Glovebox ($O_2$ and $H_2O$ < 0.5 ppm) and heated at 110 °C for 10 minutes to remove residual adsorbates. For all 2D materials examined in this study, commercially purchased bulk crystals (3R-$MoS_2$ crystals from HQ Graphene, and 2H-$MoS_2$ from 2D Semiconductors) were mechanically exfoliated onto the solvent cleaned Si/SiO$_2$ substrates inside the glovebox. Fresh MoS$_2$ bulk crystals are pressed onto the tape and cleaved once, avoiding multiple folds to minimize mechanical strain and optimize flake yield across bilayer-to-bulk thicknesses[68]. The tape is then pressed onto the heated SiO$_2$ substrate (90 °C for 35–40 s), followed by slow peeling (~25–30 s for 1 cm × 1 cm area) after cooling, yielding clean flakes across a broad thickness range. The thickness of the flakes was confirmed with optical contrast differences between the 2D material and the substrate[69], where the optical contrast with thickness trends are confirmed beforehand using atomic force microscopy.

To promote strong substrate adhesion, the samples are placed inside a tube furnace with continuous forming gas (95% Ar and 5% H$_2$) flow and annealed at 225 ºC for 2.5 hours, then left overnight to cool down to room temperature. To verify enhanced substrate adhesion from the forming gas anneal, the samples are placed into an ultrasonic acetone bath for 30 minutes, where we confirm no flakes were delaminated from the bath[53]. The final highly adhered samples are immediately placed under vacuum inside the evaporation chamber for the thin film stressor encapsulation.

A multilayer metal stressor technique is introduced in our strain engineered moiré inducing processes onto the non-twisted MoS$_2$ flakes. For the stressor encapsulation stacks, we use electron-beam evaporated Al (7 nm)/Au (20 nm)/Cr (X nm) to encapsulate the samples. The Cr thickness is varied between 25 and 125 nm, to directly tune the film force applied to the samples (see Supplementary Figure S1 for film force with Cr thickness dependence). To minimize environmental adsorbates and achieve ultra-high vacuum, the chamber is pumped overnight after sample loading, reaching a base pressure of ~1.5 × 10$^{-7}$ Torr at 23.5 °C. Finally, Al and Au layers are deposited at 0.5 Ås$^{-1}$ with an initial chamber pressure of ~10$^{-7}$ torr while the Cr layer is deposited at 0.7-1 Ås$^{-1}$. We note that the e-beam evaporated Al has been proven to mitigate damage to various 2D materials[70], thus we implement this layer first to serve as a protection layer from subsequent e-beam evaporation damage and for the subsequent etching step. Additionally, we found that directly depositing Au/Cr onto the 2D materials led to significant delamination after the etching step, likely owing to Au's strong adhesion to 2D materials[71].

In order to remove the deposited stressor layer for imaging, the samples are first placed inside a diluted $KI/I_2$ solution (Transene Gold Etchant TFA, 1:1 ratio with deionized water) for 60 seconds to remove the Au/Cr layers. The remaining Al layer on the samples is removed by soaking inside Microposit MF319 (TMAH-based photoresist developer) for 15 minutes. We implement a deionized water rinse step after each of the etchant soaks to ensure cleanliness. Finally, the samples are examined optically to confirm that the stressor layer is completely removed from the 2D flakes prior to moiré imaging.

For lithographically patterned samples, exfoliated flakes are initially uniformly encapsulated with a protective layer of e-beam evaporated Al (7 nm), before lithographic processing. This eliminates any photoresist residue from contaminating the surface since the Al and resist residue is etched away in the TMAH-based development step. Photolithography was performed using a direct-write laser photolithography system (Microtech LW405 laserwriter), with Shipley S1805 photoresist. Samples were soaked in chlorobenzene for 5 min after exposure and blown dry with $N_2$ before development with MF319. The chlorobenzene soak step is added to optimize photoresist undercut for higher quality lift-off. After the same stressor deposition process as described above, lift-off was performed using a 15-minute soak in Remover PG at 60°C.

Hyperspectral Raman imaging samples are prepared using the following steps. We used gold-assisted large area exfoliation to produce 2H bilayer $MoS_2$ with lateral dimensions of 250 μm into a 285 nm $SiO_2$ on Si substrate[23]. We start by depositing 100 nm of gold onto a bare silicon wafer (Nova Wafers Inc.) using an electron-beam evaporator (Temescal e-beam evaporator). We then spin coat polyvinylpyrrolidone (PVP) on top of the gold as an intermediate layer and we peel Au/PVP using a thermal release tape (TRT) and attach the gold surface to the freshly cleaved $MoS_2$ crystal. We peel off the

Au/PVP/TRT stack from the MoS$_2$ crystal where monolayer-multilayer MoS$_2$ are attached to the gold surface. We now attach the whole stack to a clean Si/285 nm SiO$_2$ substrate. We release the TRT by heating at 130 °C and remove the PVP by dissolving it in multiple baths of deionized water. We etch the gold using KI/I$_2$ solution leaving only monolayer-multilayer MoS$_2$ on the Si/SiO$_2$ substrate. Stressors are patterned using metal TEM grid (Ted Pella Inc.) as a shadow mask. We deposited a thin layer of MgO$_x$ at a rate of 0.5 Å/s using an electron-beam evaporator (Kurt J. Lesker Company).

**Film Force Calibration**

To extract the film force from the stressor layers stack, cleaned glass coverslips are placed alongside the samples during the deposition. The pre-stressor radius of curvature ($r_{pre}$) of the coverslips are measured along two perpendicular paths of the glass surface using a contact profilometer, then the post-deposition radius of curvature ($r_{post}$) is remeasured with the stressor films along the same paths. Employing the Stoney equation[72], we quantify the average stacked film stress ($\sigma_f$) using:

$$\sigma_f = \frac{E_s t_s^2}{6(1-v_f)t_f}\left(\frac{1}{r_{post}} - \frac{1}{r_{pre}}\right),$$

where $E_s$, $v_f$, $t_s$, $t_f$ are the coverslip's Young's modulus, Poisson's ratio for the thin film (stressor layer), and coverslip and film thicknesses, respectively. Utilizing the film stack thickness $t_f$, we measure the applied film force onto the materials $F_f = t_f \times \sigma_f$ which is used to tune the applied strain magnitude to our samples.

**Tilt-angle Dependent Scanning Electron Microscopy**

To directly visualize moiré patterns, we place the samples inside a Zeiss Auriga scanning electron microscope (SEM) as demonstrated in previous work[43]. Here, the angle between the incident beam and the various stacking order configurations will vary the amount of secondary electron emission, where one domain (AB or BA) will induce more channeling than the other. We image our stressed 3R-MoS$_2$ samples using a stage tilt of 18º, accelerating voltage of 0.5 kV, 30 μm aperture, using the secondary electron detector (SE2). To visualize H-stacking domain reconstructions, the 2H-MoS$_2$ samples are imaged at 0° tilt to preserve reconstruction lattice-specific contrast. Carbon redeposition upon SEM imaging may reduce our ability to view samples before and after stressor application[73]. This was mitigated through introducing SEM beam showering before imaging to limit the effect of carbon contamination and directly confirm that the moiré pattern derives from the stressor encapsulation and strain engineering.

**Torsional Force Microscopy**

Torsional force microscopy (TFM) measurements[74] were conducted using a Bruker Dimension Icon with Nanoscope V Controller. Using a NSC18/Pt tip, we image using a resonance frequency of ~375 kHz, a deflection setpoint of ~1 V, and a drive amplitude of ~25 mV.

**Raman Spectroscopy**

We used a Nanophoton 11 confocal Raman system to perform the Raman hyperspectral imaging. The spectra are taken using 520 nm laser source under 50x magnification with a grating of 2400 lines/mm. We used a line-scan mode with a step size of 0.39 μm with a laser power of 0.88 mW/line and an integration time of 5s.

**Molecular Statics (MS) Simulation**

Molecular statics (MS) simulations were carried out using the LAMMPS package, incorporating machine-learned interatomic potentials generated via the DeepMD toolkit[75–77]. These simulations were designed to employ hetero-strain in a vertically stacked 3R-MoS$_2$ flake of 100 nm length. The initial multilayer configuration was prepared in the 3R-stacked arrangement, with atomic positions pre-relaxed using DFT-based structural optimization. To mimic an experimental stressor-substrate setup, we incrementally imposed uniform biaxial strain on the top MoS$_2$ layer by applying a linear distribution of displacement along both in-plane directions. The topmost layer was strained and kept fixed after each step, while the other layers were allowed to fully relax during energy minimization, done using the conjugate gradient method. The uniform biaxial strain is defined as $\varepsilon_{biaxial} = \sqrt{\varepsilon_x^2 + \varepsilon_y^2}$, where $\varepsilon_x$ and $\varepsilon_y$ are engineering strains along the x and y directions respectively[20,50]. This quasi-static approach at T=0 K allowed the system to relax its interlayer interactions and atomic positions, enabling us to capture realistic reconfigurations triggered by van der Waals forces[78]. The simulation cell boundaries were kept free in all directions to mimic experimental conditions with unconstrained flakes, and a vacuum region of 40 Å was included in the out-of-plane (z) direction to prevent interactions with periodic images. Also, the free edges were passivated with hydrogen atoms to minimize artificial edge effects due to dangling bonds. Structural analysis was conducted using the OVITO visualization toolkit[79]. We employed its built-in atomic strain computation module to extract average strain transfer characteristics from the top to the bottom layer and to visualize local strain distributions throughout the flake. This analysis was key to identifying the emergence of strain-induced moiré domains and understanding their spatial evolution under externally applied mechanical deformation.

**Density Functional Theory (DFT) calculations**

The initial real-space unit cell of multilayer 3R-stacked MoS$_2$ was constructed using the commercial simulation platform QuantumATK[80]. This configuration served as the input for subsequent first-principles calculations to achieve full structural relaxation. These calculations were performed within the Quantum ESPRESSO open-source package. Exchange-correlation interactions were treated using the generalized gradient approximation (GGA), employing the Perdew–Burke–Ernzerhof (PBE) functional[81]. Ion–electron interactions were described with ultrasoft pseudopotentials, which offer an efficient balance between computational speed and accuracy. To accurately model interlayer interactions, especially relevant for vdW layered materials like MoS$_2$, we incorporated Grimme DFT-D2 semi-empirical dispersion correction[82]. This approach has been widely validated in previous studies of similar 2D systems. The electronic wavefunctions were expanded using a plane-wave basis set with an energy cutoff of 40 Ry (~544 eV), while the charge density cutoff was set to 450 Ry (~6123 eV). The energy cutoff ensures the inclusion of sufficient plane waves to resolve the electronic states, while the charge density cutoff governs the resolution of the electron density, both of which are critical for convergence and reliable force calculations.

Sampling of the Brillouin zone was carried out using a Monkhorst–Pack k-point mesh of 12 × 12 × 1, which provides adequate resolution for layered 2D systems with in-plane periodicity and out-of-plane vacuum separation. The system was relaxed until the maximum residual force on each atom was less than 0.01 eV/Å. A 25 Å vacuum layer was applied along the out-of-plane (z) direction to eliminate spurious interactions between periodic images. The relaxed in-plane lattice constants for the 3R-MoS$_2$ structure were found to be a = b = 0.319 nm, consistent with literature-reported values. This optimized unit cell served as the foundation for building large-scale atomistic models used in subsequent molecular statics simulations.

**Theoretical Expression for Domain Density Calculations**

The length of moire superlattice $L_M$ is given by, $L_M = \frac{a}{\sqrt{\delta^2+\theta^2}}$, where $a$ is the lattice constant, $\delta$ is the lattice mismatch between layers, and $\theta$ is the twist angle. For moire superlattices created with very small $\theta$ (or no twist), $L_M = \frac{a}{\delta}$. For $L_M$ due to heterostrain, the superlattice wavelength is primarily governed by the intrinsic lattice mismatch between the layers caused by heterostrain, and is not sensitive to the uncertainty of small twist angles. Moreover, the moire density ($n_0$) of the obtained moire superlattices is given by $n_0 = \frac{1}{\frac{\sqrt{3}}{4}L_M^2} = \frac{4\delta^2}{\sqrt{3}a^2}$ [8,9]. For net biaxial hetero-strain ($\varepsilon$) being applied, $\delta = \frac{a'-a}{a} = \frac{a\left(1+\frac{\varepsilon}{\sqrt{2}}\right)-a}{a} = \frac{\varepsilon}{\sqrt{2}}$. This equation can be further corrected by applying the correction factors based on strain transfer efficiency ($\eta_i$) to the second and third subsequent layers (for top layer, $\eta_1 = 1$). The corrected lattice mismatch at each interface ($\delta_{i|i+1}$) can be derived as $\delta_{i|i+1} = \frac{(\eta_i - \eta_{i+1})\varepsilon}{\sqrt{2} + (\eta_{i+1})\varepsilon}$, and the corrected moiré density be expressed as $n_{0\,(i|i+1)} = \frac{4(\delta_{i|i+1})^2}{\sqrt{3}a^2}$. We obtained strain transfer efficiency to each layer by calculating the overall change in length, i.e., nominal strain ($\frac{\Delta L_i}{L_i}$) of the layers at each strain increment during MS simulations (Supplementary Table S1).

**FEA Models of Strain Distribution Across Different Stressor Geometries**

We used the commercial FEA package ABAQUS Standard solver. Material properties are summarized in Supplementary Table S2. A 20× 20 μm wide and 0.1μm thick rectangular solid was chosen for substrate (bulk MoS$_2$). For monolayer MoS$_2$, we choose a slightly smaller size (19.5× 19.5 μm) such that MoS$_2$ under strain doesn't extend outside the substrate. To address the high aspect ratio of 2D materials we assume the thickness of the 2D layer to be 1 nm and use scaling law ($E_{bulk}t_{2D} = E_{model}t_{model}$) to adjust the elastic modulus. The size of the stressors are chosen such that they are substantially inside the 2D layer (i.e., 6 μm width for square) so that the free boundaries of the monolayer don't affect the strain field. All simulations are for stressors of thickness 60 nm. We do not model the interlayer dynamics of a bilayer and instead assume a monolayer on a bulk MoS$_2$ substrate. The interfacial properties are modeled as surface-to-surface interaction between the MgO$_x$-bottom/MoS$_2$-top and MoS$_2$-bottom/MoS$_2$ bulk surfaces using cohesive behavior with specified damage evolution properties. Here, the interfacial traction coefficients, $K_t$ and $K_b$, and damage initiation threshold, $\delta_d$, are defined. The simulations in the text use a $K_t$ of $10^4$ MPa/μm and $K_b$ value of 50 MPa/μm, while ensuring that $K_t \gg K_b$. $\delta_d$ value in simulation range from 20-40 nm. Small sliding capability is included to enable the contacting surfaces to undergo small sliding relative to each other. To determine the interfacial parameters, we observed that the strain decay lengths as measured by the density of soliton domains in these structures are in the order of few microns and thereby used parametric analysis from previous studies[23] to estimate the interfacial traction coefficient to be 50 MPa/μm. In this model, the mesh size is largely determined by the thickness of the thinnest material, which is 0.001 μm for MoS$_2$. We used quadratic tetragonal 3D elements with a mesh size ranging from 0.02- 0.12 μm. The built-in stress of MgO$_x$ is defined as a predefined stress field of −1.0 GPa. The bottom nodes of the bulk MoS$_2$ are constrained against all degrees of freedom, i.e., displacement and rotation.